\documentclass{article}
\usepackage[utf8]{inputenc}
\usepackage{authblk}
\usepackage{comment}
\usepackage{braket}
\usepackage{amsmath}
\usepackage{float}
\usepackage{tikz}
\usetikzlibrary{quantikz}
\usetikzlibrary{positioning,shadows,shapes,arrows}
\usepackage{dsfont}
\usepackage{geometry}
\usepackage{xcolor}
\usepackage{graphicx}
\usepackage{dcolumn}
\usepackage{bm}
\usepackage{hyperref}
\usepackage{microtype}
\usetikzlibrary{arrows}
\usetikzlibrary{arrows.meta}
\usetikzlibrary{shapes,arrows}
\usepackage{amsmath,bm,times}
\usepackage{tikz}
\usetikzlibrary{arrows.meta}

\usepackage{physics}
\usepackage{xcolor}
\usepackage[linesnumbered,ruled,vlined,boxed]{algorithm2e}
\usepackage[sorting=none]{biblatex}

\usepackage[capitalise,compress]{cleveref}\crefname{section}{Sec.}{Secs.}
\Crefname{section}{Section}{Sections}
\crefrangelabelformat{equation}{\textup{(#3#1#4)}--\textup{(#5#2#6)}}
\newtheorem{thm}{Theorem}

\crefname{cor}{Cor.}{Cors.}

\title{Classical Algorithm for the Mean Value problem over Short-Time Hamiltonian Evolutions}
\author[1]{Reyhaneh Aghaei Saem}
\author[1]{Ali Hamed Moosavian}
\affil[1]{Phanous}
\date{}
\begin{document}
\maketitle

\begin{abstract}
    Simulating physical systems has been an important application of classical and quantum computers. In this article we present an efficient classical algorithm for simulating time-dependent quantum mechanical Hamiltonians over constant periods of time. The algorithm presented here computes the mean value of an observable over the output state of such short-time Hamiltonian evolutions. In proving the performance of this algorithm we use Lieb-Robinson type bounds to limit the evolution of local operators within a lightcone. This allows us to divide the task of simulating a large quantum system into smaller systems that can be handled on normal classical computers.
\end{abstract}

\section{Introduction}

Understanding physical systems with quantum mechanical interactions have been an increasingly challenging and important task in the past century. Because nature is inherently quantum, many naturally occurring or artificially implemented phenomena cannot be explained without quantum mechanics. Some early examples include understanding atoms \cite{pauli_uber_1926}, diatomic molecules \cite{mensing_rotations_1926}, the Meissner effect in superconductors \cite{london_electromagnetic_1935} and black-body radiation \cite{planck_ueber_1901}. With the advent of better theoretical and computational tools, the list of phenomena that have been shown to require quantum mechanics has grown enormously in the past few decades. To name a few, some of the more notable examples include photosynthesis \cite{hayes_engineering_2013}, topological ordered phases such as fractional quantum Hall effect \cite{zaletel_topological_2013}, isomerization of diazene \cite{Arute2020} and non-Abelian lattice gauge theories \cite{Banerjee2012}. Indeed, because of the challenging nature of simulating quantum systems on classical computers, one of the earliest motivations for quantum computers has been to use them to study other quantum systems \cite{Feynman1982}.

Besides the obvious practical applications, the problem of simulating quantum systems has some critical complexity theory value too. On the one hand, it is known that the problem of simulating several quantum systems belongs to the Bounded-error Quantum Polynomial-time complete (BQP-complete) class \cite{Jordan2018,bao_universal_2015,chen_heun_2012}. Also, there are specific problems where the complexity class of simulating them varies as the runtime increases, and exhibit a dynamical phase transition \cite{Deshpande2018,ehrenberg_simulation_2022}.

In this paper, we present a classical algorithm for computing the expectation value of an observable that can be written as a tensor product of local operators acting on each qubit. In the literature this problem is sometimes called the quantum mean value problem \cite{Bravyi2021} (not to be confused with a similarly named open problem in mathematics \cite{smale_fundamental_1981}). Our algorithm evaluates the mean value of an operator where the state is generated by evolving a product state under a geometrically local time-dependent Hamiltonian for a short period of time. 
The setup of our problem is motivated by the state of current quantum technologies. On one hand, analog simulators lack fault-tolerance, which means they have a limited decoherence time, and on the other hand, we still do not have access to fault-tolerant universal quantum computers either.  
 
Suppose that we have a bounded-norm time-dependent Hamiltonian $ \vb{H(t)} $ that acts on $n$-qubits. The initial state of the system is assumed to be a product state, typically $\ket{\psi(0)}= \ket{0}^{\otimes n}$. The quantum mean value problem wants to compute the expectation value of an operator $\vb{O}$ with respect to $\ket{\psi(T)}$. The expectation value is represented with $\mu$:
\begin{equation}
    \mu \equiv \ev{\vb{O}}{\psi(T)}\, .
\end{equation}
We consider geometrically local time-dependent Hamiltonians that are defined on 2D or 3D lattices. 1D systems with gapped Hamiltonians have been thoroughly studied before \cite{Somma2015,Schollwock2011a}.

The unitary evolution operator corresponding to this Hamiltonian can be written as:
\begin{equation}
    \vb{U}(t) = \mathbf{T} :\qty[ \exp\qty(- \frac{i}{\hbar} \int_{0}^{t} \vb{H}(t^\prime) dt^\prime)]\, ,
    \label{2}
\end{equation}
where $\mathbf{T}:\qty[.]$ means the operators respect time ordering.
Assuming the observable can be written as the tensor product of single-qubit Hermitian operators $\vb{O}_j$ acting on individual sites, $\vb{O} = \vb{O}_1 \otimes \vb{O}_2 \otimes \dots \otimes \vb{O}_n$, we can write the mean value of the observable $\vb{O}$ as:
\begin{equation}
    \mu = \langle 0^n | \; \vb{U^\dagger}(T) \vb{O}_1 \otimes \vb{O}_2 \otimes \dots \otimes \vb{O}_n \vb{U}(T) \; | 0^n \rangle\, .
    \label{3}
\end{equation}

In this paper, we introduce a classical algorithm that can approximate the mean value problem for time-dependent Hamiltonians with an additive error. The structure of the rest of this paper is as follows. In \cref{sec: ClassicalAlgorithm} we give an overview of the algorithm. \Cref{sec:LocalUnitaries} explains how the dynamics of each operator is approximately restricted to within a lightcone. In \cref{sec: ClassicalSimulationOfTimeDepUnitaries} we analyze the problem of classically calculating the local unitary operators and compare different numerical algorithms for doing the task and in \cref{sec:Conclusions} we conclude. 

\section{Algorithm}\label{sec: ClassicalAlgorithm}
We provide a classical algorithm for approximating $\mu$ within an additive error, $\delta$, for the special case of the mean value problem where the time-dependent Hamiltonian is geometrically local in 2D or 3D.

Conceptually, the algorithm can be broken into two parts. First we use a Lieb-Robinson bound \cite{nachtergaele_propagation_2006, Moosavian2021, sweke_liebrobinson_2019} to limit the unitary evolution corresponding to each qubit within a lightcone. This allows us to classically compute the unitary operator. Then we follow the steps in \cite{Bravyi2021} to divide the lattice into pseudo 1D slices, that can be efficiently simulated either using Matrix Product State algorithms \cite{vidal_efficient_2003,yoran_classical_2006,jozsa_simulation_2006} (for a nice review of these algorithms see \cite{Schollwock2011a}) or the algorithm ascribed to \cite{Bravyi2021}.
\begin{thm}
Let $\vb{H}\qty(t)$ be a bounded-norm time-dependent lattice Hamiltonian that acts on $n$-qubits. Suppose the observable $\vb{O}$ is a tensor product of operators , $\vb{O}= \vb{O}_1\otimes \vb{O}_2 \otimes \dots \otimes \vb{O}_n$. For constant evolution times, there exists a classical algorithm that estimates $\mu$ within an additive error $ \delta $,
\begin{equation}
    | \tilde{\mu} - \mu | \leq \delta \, .
\end{equation}
\end{thm}
The error $\delta$ includes three different parts. The first contribution comes from the Lieb-Robinson bound where we used it to approximately limit the evolution within the lightcone of each qubit. Second, numerical methods such as trotterization are used to calculate the local unitary evolutions; these methods are not exact and incur some errors. The third part comes from the additive error in simulating a constant depth quantum circuit \cite{Bravyi2021}. 
In \cite{Bravyi2021}, they provide a classical algorithm for constant depth circuits in 2D or 3D which can approximate the mean value to an additive error. In our work, the short-time evolution of a geometrically local Hamiltonian which is limited by the Lieb Robinson bound is comparable with the shallow quantum circuit with a constant depth. 

Suppose that the error of localizing the time evolution of the Hamiltonian, classical simulation of time-dependent unitaries and simulating shallow circuits are given by $\varepsilon^{LR}$, $\varepsilon^{CS}$ and $\varepsilon^{SSC}$ respectively. Then the total error is as follows:
\begin{equation}
    |\tilde{\mu} - \mu| \leq \varepsilon^{LR} + \varepsilon^{CS} + \varepsilon^{SSC} = \delta\, .
\end{equation}

In \cref{eq: Lieb-Robinson}, we will see how the first error is dictated by the simulation time, the lightcone radius and the Hamiltonian. One should note that $\varepsilon^{LR}$ is bounded by $\varepsilon^{LR}\le n \varepsilon^{(LR)}(L,T)$, where $ \varepsilon^{(LR)}(L,T) $ is defined later in \cref{eq: Lieb-Robinson} as the Lieb Robinson error for a single site. \Cref{sec: ClassicalSimulationOfTimeDepUnitaries} derives the dependency of the second error term on the simulation parameters. 

\begin{algorithm} 
    \DontPrintSemicolon
    \SetKwData{Left}{left}\SetKwData{This}{this}\SetKwData{Up}{up}
    \SetKwFunction{Union}{Union}\SetKwFunction{FindCompress}{FindCompress}
    \SetKwInOut{Input}{input}\SetKwInOut{Output}{output}
    \Input{a $\sqrt{n}\times \sqrt{n}$ lattice of qubits, a geometrically local time-dependent Hamiltonian $\vb{H}(t)$, the operator $\vb{O}= \vb{O}_1\otimes \vb{O}_2 \otimes \dots \otimes \vb{O}_n$ where each $\norm{\vb{O}_j}\le 1$, an upper bound for error $\delta$, a simulation time $T$.}
    \Output{an approximation for $\mu=\ev{\vb{O}}{\psi(T)}$, where $\ket{\psi(0)}=\ket{0^{\otimes n}}$ and $i\hbar \frac{d}{dt}\ket{\psi(t)}=\vb{H}(t)\ket{\psi(t)}\,$}
    \BlankLine
    Initialization: \;
    Use $T$ and $\delta$ to calculate the lightcone radius, $L$, from the Lieb-Robinson bound. \;
    As in \cref{fig:algorithm} partition the lattice into $4L\times\sqrt n$ strips twice, let us call each set $\qty{A_i}_i$ and $\qty{B_i}_i$. 
    Also, define the sets $\qty{A_i^0}_i$ and $\qty{B_i^0}_i$ as the central part of the strips. \;
    Also, group sites into $2L\times 2L$ super-sites to form a coarse-grained lattice.\;
    Calculations: \;
    \For{$A_i \in \qty{A_i}_i$ \label{for loop}}{
    initialize an MPS.\;
    \For{{$\vb{O}_j$} in {$A_i^0$}}{
    Use a classical ODE solver to calculate the local unitary operator $U_j$ corresponding to $\vb{O}_j$ which includes the terms that are in the lightcone of the $j$th qubit.\;
    Transform $\vb{O}_jU_j$ into a Matrix Product Operator.\;
    Add the necessary sites from the current super-site to the active memory and apply the MPO on it. \;
    Measure any sites that will no longer be needed. \;
    }
    Let us call the (not normalized) MPS outcome $\ket{\Psi_{A_i}(T)}$ .\;
    }
    Repeat the \cref{for loop} loop for $ B_j $s.\;
    \Return{$\tilde{\mu} = \qty\big(\bigotimes_j \bra*{\Psi_{B_j}(T)})\qty\big(\bigotimes_i \ket{\Psi_{A_i}(T)})$}

    \caption{High level overview of the algorithm} \label{alg:Overview2d}
\end{algorithm}

The runtime of classical simulation of the time-dependent unitaries is related to the complexity of matrix multiplication. We can find the lightcone radius $L$ from \Cref{sec:LocalUnitaries} and then find the number of qubits $m$ in each lightcone which is $\mathcal{O}\qty(L^2)$. For most practical cases, the fastest matrix multiplication algorithm is the famous Strasson algorithm with asymptotic complexity of $\mathcal{O}\qty(\qty(2^m)^{\log_2 7})$ \cite{strassen_gaussian_1969}, however, the best known asymptotic complexity for matrix multiplication is $\mathcal{O}\qty(\qty(2^m)^{2.373})$ \cite{DBLP:journals/corr/abs-2010-05846}. For most physical Hamiltonians where we have translational symmetry in the bulk, we only need to calculate the unitary evolution matrix for each geometrical configuration of the sites once. This means that the complexity of this part of the algorithm would be $\mathcal{O}\qty(2^{2.373m})$. But for generic Hamiltonians the number of configurations could grow linearly with the system size and the complexity would be $\mathcal{O}\qty(n 2^{2.373m})$. 

For 2D and 3D systems with $n$ qubits, the runtime of simulating a shallow circuit is related to the last error term with $\mathcal{O}\qty(nL^2 2^{6L^2}/ (\varepsilon^{SSC})^2)$ and $\mathcal{O}\qty(nL^3 2^{6L^2n^{1/3}}/ (\varepsilon^{SSC})^2)$ respectively \cite{Bravyi2021}. We have provided a high-level overview of the 2D algorithm in \cref{alg:Overview2d}. Consequently the general total complexity of the algorithm for 2D and 3D systems is $\mathcal{O}\qty(n 2^{2.373m} + nL^2 2^{6L^2}/ (\varepsilon^{SSC})^2)$ and $\mathcal{O}\qty(n 2^{2.373m} + nL^3 2^{6L^2n^{1/3}}/ (\varepsilon^{SSC})^2)$ respectively.

\begin{figure}
    \includegraphics[width=\textwidth]{"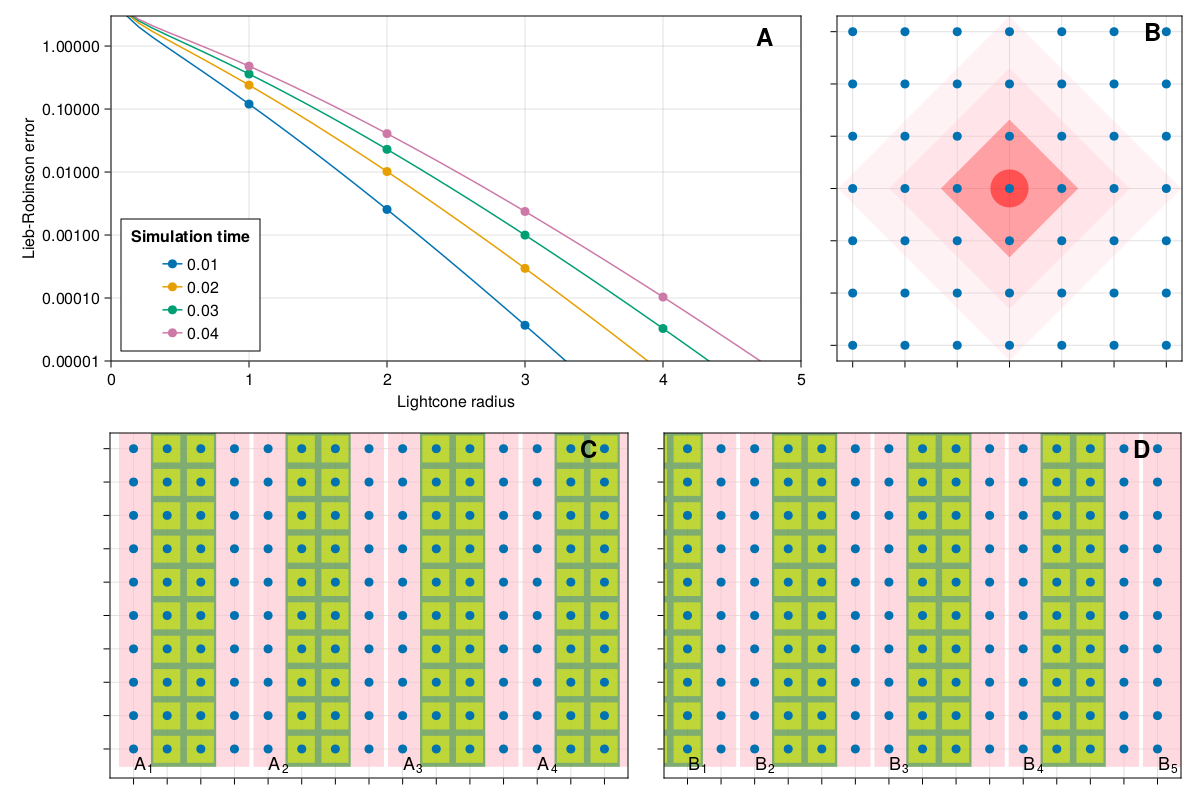"}
    \caption{\bf{A:} \normalfont It shows the relationship between simulation time and the lightcone radius. \bf{B:}\normalfont shows the sites inside the lightcone for  various lightcone radii. \bf{C and D:}\normalfont show the set of strips $\qty{A_i}_i$ and $\qty{B_i}_i$ as well as the $\qty{A_i^0}_i$ and $\qty{B_i^0}_i$ sets on a 2D grid. The yellow squares represent the $\vb{O}_i$ operators that are to be applied on either set of strips. \label{fig:algorithm}}
\end{figure}

\section{Local Unitary Operators}\label{sec:LocalUnitaries}

According to \cite{Moosavian2021}, we know that a local operator $ \vb{O}_A $ which is defined inside a region $A$, remains local after a short-time evolution under a local Hamiltonian. Suppose that the Hamiltonian has the form 
\begin{equation} \vb{H}(t) = \sum_{e} u_e(t) \vb{h}_e\, ,\end{equation}
where $\vb{h}_e$ acts non-trivially only on the two vertices of edge $e$ of the graph $G$. Suppose that $ \norm{ u_e(t) \vb{h}_e} \leq g $ for $0\le t\le T$ and the maximum degree of the graph to be $ \Delta $. The Hamiltonian for terms in region $A$ and the set of vertices in the L-boundary of it has the form
\begin{equation}
     \vb{H}_A(t) = \sum_{e \subset A \cup \partial_L(A) } u_e(t) \vb{h}_e\, .
    \label{local_Hamiltonian}
\end{equation}
The time evolution operator of this local Hamiltonian acts non-trivially only on the region $A$ and its $L$-boundary. The time evolution operator is given by
\begin{equation}
    \vb{V}(t) = \vb{T} :\qty[ \exp\qty(- \frac{i}{\hbar} \int_{0}^{t} \vb{H}_A(t^\prime) dt^\prime)] \, .
    \label{4}
\end{equation}
This is known as Dyson series and $\vb{T}:\qty[.]$ represents \begin{it}time-ordering\end{it} \cite[p.~551]{griffiths_schroeter_2018}. According to \cite{Moosavian2021} the following Lieb-Robinson bound holds:
\begin{equation}
   || \vb{U^\dagger}(T) \vb{O}_A \vb{U}(T) - \vb{V^\dagger}(T) \vb{O}_A \vb{V}(T) || \leq \varepsilon^{(LR)}(L,T)\, ,
    \label{5}
\end{equation}
where $\varepsilon^{(LR)}(L,T)$ is defined as:
\begin{equation}
    \varepsilon^{(LR)}(L,T) = \sqrt{\frac{2}{\pi}} \; |A| \; ||\vb{O}_A|| \; \exp\qty( -L(\log L - \log T - \log (4 g (\delta - 1 ))) - \frac{1}{2} \log L)\, .
    \label{eq: Lieb-Robinson}
\end{equation}
For each local operator in $\vb{O}= \vb{O}_1\otimes \vb{O}_2 \otimes \dots \otimes \vb{O}_n$, we only consider the Hamiltonian terms inside the lightcone of it. We replace the global unitary that acts on the entire system with these local unitary operators and apply the algorithm in \cref{sec: ClassicalAlgorithm} to approximate the mean value.

The only remaining problem would be to find a classical algorithm for classically calculating the local unitary operators $\vb{V}(t)$. We analyze different approaches for doing so in \cref{sec: ClassicalSimulationOfTimeDepUnitaries}.

\section{Classical Simulation of the time-dependent unitaries}\label{sec: ClassicalSimulationOfTimeDepUnitaries}
\subsection{Trotterization}

\begin{thm}
    Let $ \vb{H}_A \qty(t) $ be a bounded-norm time-dependent Hamiltonian and $ \vb{O}_A $ an observable inside region $A$. Let $ \vb{V}\qty(t) $ be the unitary evoluion of $ \vb{H}_A\qty(t) $. We can approximate the operator $\vb{V}^{\dagger}\qty(T) \vb{O}_A \vb{V}\qty(T)$ with $\vb{W}^\dagger\qty(T, N) \vb{O}_A \vb{W}\qty(T, N)$ where $ \vb{W}\qty(t, N) $ is defined as 
    \begin{equation}
        \vb{W}\qty(t, N) = \prod_{j = 1}^{N = \frac{t}{\delta t}} \exp \qty(- \frac{i}{\hbar} \delta t \, \vb{H}_A \qty(j \delta t)), \label{eq:Appr_Schr_Unitary}
    \end{equation}
    such that 
    \begin{equation}
        \norm{\vb{W}^\dagger \qty(T, N) \vb{O}_A \vb{W}\qty(T, N) - \vb{V}^\dagger\qty(T) \vb{O}_A \vb{V}\qty(T)} \leq \frac{6 T^2}{N \hbar} \norm{\vb{O}_A}  \, \norm{\vb{H}^\prime_A \qty(t^*)} = \varepsilon^{CS},  \label{eq:epsilonCS}
    \end{equation}
    where 
    \begin{equation}
        \norm{\vb{H}^\prime_A \qty(t^*)} = \max_{0 \leq t \leq T} \norm{\vb{H}^\prime_A \qty(t)}.
    \end{equation} 
\end{thm}
This is a well-known textbook result that can be found in the literature. For instance see chapter IX of \cite{Kato:1966:PTL}.

\subsection{Numerical Differential Equation Solvers}

Another approach for approximating the unitary time evolution would be to derive a differential equation from Schrodinger's equation and solve that numerically by using a suitable classical algorithm.
\begin{equation}
    \begin{aligned}
        i\hbar \frac{d}{dt}\ket{\psi(t)}&=\vb{H}(t)\ket{\psi(t)}\, , \\
        i\hbar \frac{d}{dt}\vb{U}(t)\ket{\psi(0)}&=\vb{H}(t)\vb{U}(t)\ket{\psi(0)}\, , \\
        i\hbar \frac{d}{dt}\vb{U}(t)&=\vb{H}(t)\vb{U}(t)\, .\label{Eq:Schrodinger_Unitary}
    \end{aligned}
\end{equation}

\begin{figure}
    \includegraphics[width = \textwidth]{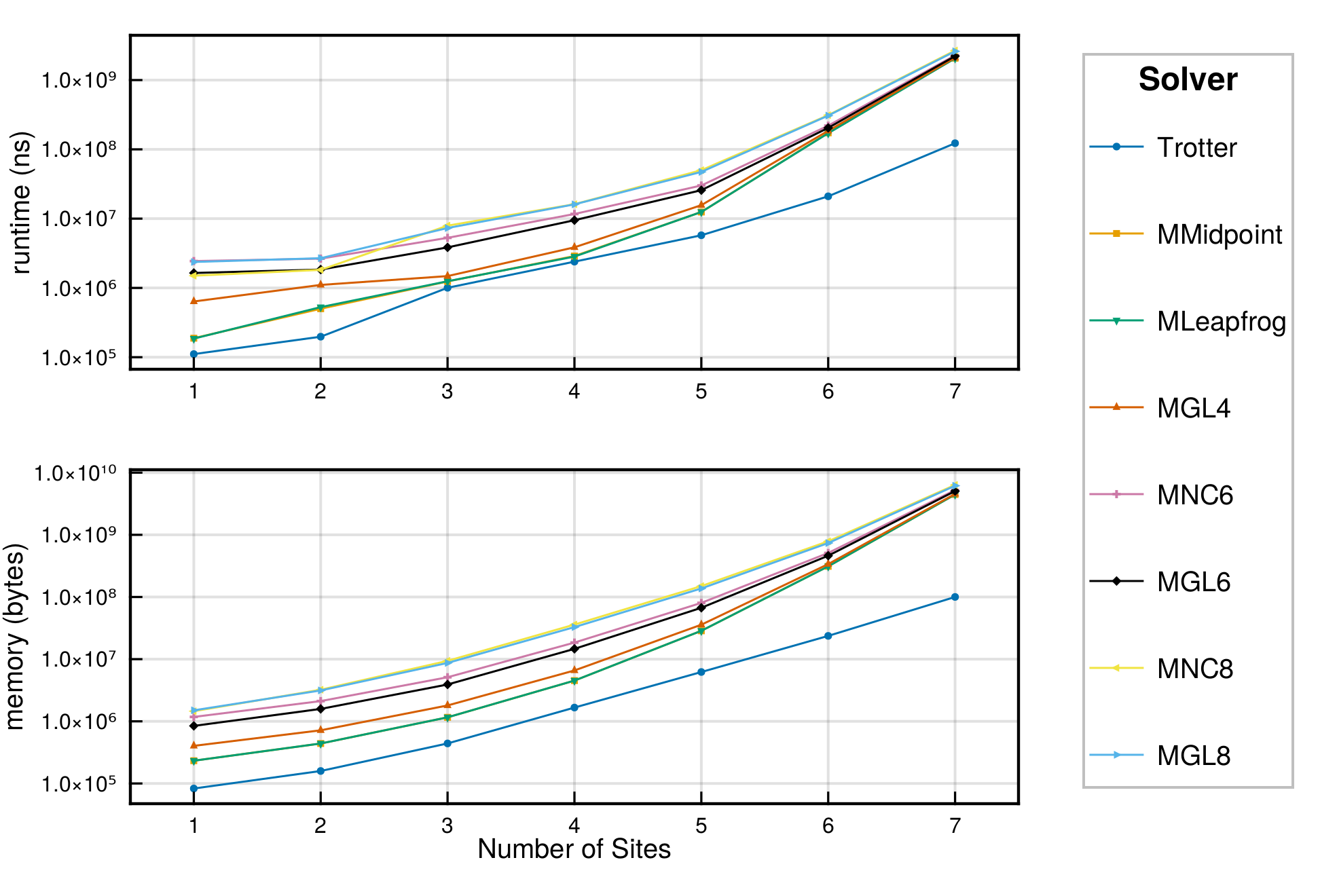}
    
    \caption{A comparison of different ODE solvers. The top figure shows average minimum time used by each ODE solver to solve \cref{Eq:Schrodinger_Unitary} for different number of qubits. The bottom plot shows the average memory used by each solver to solve the same differential equation. The benchmark was done on a personal PC and each data point was repeated at least 100 times over 20 randomly generated time-dependent Hamiltonians. All of the methods except Trotter were picked from Julia's DifferentialEquations.jl roster of state of the art ODE solvers \cite{rackauckas2017differentialequations}, and compared to the Trotter method, they consistently had at least 10 orders of magnitude less error of the form \cref{eq:epsilonCS}. The Trotter solution was also implemented in Julia, and used $N=30$.   \label{fig: ODE_comparison}}
\end{figure}

 If there are $m$ qubits inside the lightcone, the matrix representation of $\vb{U}$ and $\vb{H}$ will be $2^m\times 2^m$ operators and \cref{Eq:Schrodinger_Unitary} will constitute a set of Ordinary Differential Equations (ODEs). The upside of using a classical ODE solver is that they can consistently attain much lower errors than what is possible from \cref{eq:Appr_Schr_Unitary} or a higher order Suzuki-Trotter solution \cite{Suzuki1976,Suzuki1991} that is fine-tuned for a time-dependent problem \cite{Hatano2005,wiebe_higher_2010}. The downside is that they are typically orders of magnitude slower and require more memory than the straight forward trotterization as in \cref{eq:Appr_Schr_Unitary}.  Assuming our lightcone is small enough, many numerical ODE solvers will be able to handle \cref{Eq:Schrodinger_Unitary}. See \cref{fig: ODE_comparison} for a comparison between multiple different ODE solvers.

\section{Conclusions}\label{sec:Conclusions}
To conclude we have provided a classical algorithm for the mean value problem on outcomes of short time dependent Hamiltonian evolutions. These mean values are typically used in other algorithms such as the Variational Quantum Algorithm \cite{cerezo_variational_2021}. 

The quantum mean value problem, almost by definition, belongs to the BQP-complete class for polynomial times. So, naturally we do not expect to be able to solve the polynomial time problem on a classical computer efficiently. Nonetheless, an important open question would be to find the minimum simulation time for getting a quantum speedup. The current work shows us that in order to benefit from the quantum speedup we need simulation times that are at least greater than constant \cite{babbush_focus_2021}. This is in accordance with a wide variety of other results that target specific problems too \cite{Bravyi2021,Bravyi2020,Deshpande2018, Farhi2020, Farhi2020a,Moosavian2021}.
In general the problem of mapping out the entire dynamical complexity phase diagram is theoretically interesting.

Another direction for future research would be to improve or generalize the current algorithm. If the algorithm cannot be further improved or generalized, then proving these limitations is another open question.

\section{Acknowledgements} \label{sec:Acknowledgements}
We would like to thank Salman Beigi for overseeing this project and commenting on the draft. We also thank Leila Taghavi and Erfan Abedi for helpful discussions.

\printbibliography
\end{document}